\documentstyle[12pt,a4wide,epsfig,amssymb,axodraw]{article}

\def\gsim{{~\raise.15em\hbox{$>$}\kern-.85em
          \lower.35em\hbox{$\sim$}~}}
\def\lsim{{~\raise.15em\hbox{$<$}\kern-.85em
          \lower.35em\hbox{$\sim$}~}}

\title{
\vspace*{-2.0cm}
\begin{flushright}
{\normalsize CPT-2001/P.4261}
\end{flushright}
\vspace{1.4cm}
{\Large \bf Leptogenesis at the TeV scale
\unboldmath} 
\vspace*{0.8cm}
}
\author{ 
T.~Hambye\footnote{\small E-mail: hambye@cpt.univ-mrs.fr}
\\[0.5cm]
\normalsize  {\it Centre de Physique Th\'eorique, 
Luminy Case 907, F-13288 Marseille Cedex 09, France}\\[1mm] 
}
\date{}
\begin{document}
\maketitle
\thispagestyle{empty}
\vspace*{+5.2cm}
\begin{abstract}
We present a general description of the problems encountered when attempting 
to build a simple model of leptogenesis and hence of baryogenesis 
at an energy scale as low as 1-10 TeV.
We consider three possible lepton asymmetry enhancement mechanisms 
in the out-of-equilibrium decay scenario, emphasizing 
the three body decay mechanism as most natural. A
new model based on the three body decays of 
right-handed neutrinos is proposed. It naturally 
allows both leptogenesis and neutrino mass
generation at low scale. Also 
discussed is the possibility of inducing leptogenesis
at low scale in existing neutrino mass models: Fukugita-Yanagida model,
Higgs triplet model, Zee model and 
models with R-parity violation.
\end{abstract}
\vspace*{\fill}
%
%%%%%%%%%%%%%%%%%%%%%%%%%%%%%%%%%%%%%%%%%%%%%%%%%%%%%%%%%%%%%%%%%%%%%%%
\newpage
%%%%%%%%%%%%%%%%%%%%%%%%%%%%%%%%%%%%%%%%%%%%%%%%%%%%%%%%%%%%%%%%%%%%%%%
%
\section{Introduction}
The baryon asymmetry of the universe is usually expressed in terms of the
ratio of the baryon density $n_B$ to the entropy density $s$ of the universe.
From nucleosynthesis constraints this ratio is 
determined to be \cite{bu}:
\begin{equation}
\frac{n_B}{s}\simeq (6-10) \cdot 10^{-11} \label{nbs}
\end{equation}
in good agreement with the latest results from Cosmic Microwave 
Background data \cite{cmb}.
To explain this asymmetry the electroweak baryogenesis mechanism is
certainly the most attractive, especially because it is testable.
However, in the Standard Model this mechanism is now 
excluded \cite{spra,dol}.
In the minimal supersymmetric model 
this mechanism could be operative but only if a certain number 
of rather restrictive conditions are satisfied \cite{quiros}.
In this context 
it is important to look for other alternatives. Beside
the electroweak baryogenesis mechanism in the context of more 
complicated models,
leptogenesis [6-15]
%\cite{fy,lpy,luty,akms,fps,bp,pil,ht,mr,moh} 
is probably the most attractive and simple alternative. 
It is based on a two step process. First, at a certain 
temperature in the thermal evolution of the universe, a lepton asymmetry is
produced. Secondly, once this asymmetry has been produced, it is
partly converted to a baryon asymmetry by the sphaleron 
processes \cite{klink,sph}
which are very fast $B+L$ violating 
processes in thermal equilibrium at temperature above
$\sim 100$-200~GeV \cite{spra,latt,moore}. 
Probably the most attractive model of leptogenesis
is the one of Fukugita-Yanagida \cite{fy} based 
on heavy right-handed neutrinos. 
The Heavy Higgs triplet model of Ma-Sarkar \cite{triplet,hms3} 
is also very interesting and simple.
Both models have the very attractive feature that the interactions
at the origin of the lepton asymmetry also induce naturally small neutrino 
masses via the seesaw mechanism. However,
these models of leptogenesis and neutrino masses have
an intrinsic problem for phenomenology: their lack of testability.
In those models the natural scale of the interactions at the origin
of both leptogenesis and neutrino masses (in agreement 
with atmospheric and solar neutrino data)
lies from $10^{10}$ GeV to $10^{15}$ GeV. 
It is thus particularly interesting to look for alternatives at
lower energy scales of order 1-10 TeV which would be directly testable in a 
relatively near future.
In this paper the question of
how to build a successful leptogenesis mechanism
at the 1-10 TeV scale is addressed in detail.

In section 2 we begin with introducing the usual Fukugida-Yanagida 
leptogenesis model. This 
is useful to illustrate the various problems encountered 
when we attempt to build at low scale 
a leptogenesis mechanism in the out-of-equilibrium 
decay scenario.\footnote{The possibility of generating baryogenesis at
low energy from other scenarios such as the Affleck-Dine scenario \cite{ad} 
will not be considered here. Similarly the possibility of generating 
leptogenesis by assuming large extra dimensions will also not be considered.}
These problems will be listed in section 3. 
In section 4 we consider three possible mechanisms for the enhancement 
of the asymmetry,
which could be used to remedy these problems.
We emphasize the fact that 3 body decays
can lead naturally to a sufficiently large lepton asymmetry without the need for
unnatural hierarchies of couplings
or finely tuned mass degeneracies (as the other enhancement 
mechanisms do). At the end of section 4 we briefly review
whether the three enhancement mechanisms introduced may 
lead to a successful leptogenesis mechanism at low energy in the 
framework of existing neutrino mass models: 
the Fukugita-Yanagida model, the Higgs triplet model, the Zee model
and the models with R-parity violation. 
We stress that, except for the debatable "mass degeneracy" enhancement 
mechanism
in the Fukugita-Yanagida model, it is very difficult if
not impossible to build
a successful low scale mechanism in these frameworks.
After this general description of the problems and mechanisms,
a minimal model is presented in section 5 which, 
using the 3 body decay
mechanism, 
avoids all these problems.
It is based on the decay of
heavy right-handed Majorana neutrinos and the existence of 
2 scalar charged singlets. This model allows for a consistent
generation of both baryogenesis and neutrino masses at the 1-10 TeV scale.
A reader familiar with this topics may choose to skip the general 
discussions of sections 2-4 and 
go directly to the discussion on three body decays in section 4 and then to 
section 5 where
the original model is presented.
Our conclusions are contained in Section~6. 
%
%%%%%%%%%%%%%%%%%%%%%%%%%%%%%%%%%%%%%%%%%%%%%%%%%%%%%%%%%%%%%%%%%%%%
%
\section{The heavy right-handed neutrino mechanism \label{CKM}}

We begin with introducing briefly the Fukugita-Yanagida model \cite{fy} 
which is useful to
illustrate in the next sections the various problems encountered when
trying to build a simple leptogenesis model at the 1-10 TeV scale.
This model is based on 
the existence of 3 
self-conjugate (i.e. Majorana) neutrinos $N_{i}$ which are singlets under
$SU(2)_L$. These 
neutrinos are expected to have Majorana
masses $M_{N_i}$ much larger 
than the electroweak scale since these masses are not protected by the
$SU(2)_L \times U(1)$ gauge symmetry. They naturally couple to
one lepton doublet and the standard model scalar doublet via the usual
Yukawa interactions. The lagrangian of the model is therefore:\\
\begin{equation}
{\cal L}_{FY}={\cal L}_{SM} + \bar{\psi}_{Ri} i \partial \!\!\!/ \psi_{Ri} - 
\frac{M_{N_{i}}}{2}( \bar{\psi}_{Ri} \psi_{Ri}^c + h.c.)
+ (h_{ij} \bar{L}_j \psi_{Ri} \Phi + h.c.) \, , \label{Lfy}
\end{equation}
where the $\psi_{Ri}$ are the two component Majorana spinors which in terms 
of the 4-components  self-conjugated Majorana spinor $N_{i}$ 
are given by $N_{i}=\psi_{Ri} + \psi_{Ri}^c$ (with $\psi_{Ri}=
\frac{1}{2}(1 +\gamma_5) N_i$
and $\psi_{R}^c = P_L C \bar{\psi}^T$).
$L_i=(\nu_{Li} \, \, l_{Li})^T $ and $\Phi=( \phi^0 \, \, \phi^-)^T$.
The Yukawa interactions in Eq.~(\ref{Lfy}) 
are at the origin of seesaw induced neutrino 
masses \cite{seesaw}:
\begin{equation}
(m_\nu)_{ij}= (h)_{ik} M_{N_{k}}^{-1} (h^\dagger)_{kj}\frac{v^2}{2} \, ,
\label{mnufy}
\end{equation}
with $v=\sqrt{2} \langle \phi^0 \rangle =246$~GeV.
As is well-known these masses are naturally small 
if the masses of the 
right-handed neutrinos are heavy.
To have a neutrino mass of order the SuperKamiokande bound
($\sim0.1$~eV) with for example Yukawa couplings $h$ of 
order $10^{-2}$ we need $M_{N_{i}} \sim 10^{10}-10^{11}$~GeV.

Beside inducing the $\nu$ masses, the Yukawa couplings are also at
the origin of the right-handed neutrino decays
$N_{k} \rightarrow L_j + \Phi^\ast$. The decay width is:
\begin{equation}
\Gamma_{N_{k}}=\frac{1}{8 \pi} \sum_j |h_{kj}|^2 M_{N_{k}} \, .
\label{gamfy}
\end{equation}
From these decays a lepton asymmetry can be created. The Yukawa couplings
$h_{ij}$ provide the source of 
CP-violation which is necessary for the creation of the asymmetry. 
In the decay, this CP-violation can manifest 
itself only at one loop level, where it is associated with the 
imaginary part of one
loop diagrams. The lowest order non-trivial asymmetry comes from 
the interference of the tree level diagrams
with the imaginary part of the 
one loop diagrams of Fig.~1.
There are two types of one-loop diagrams: vertex diagrams
as considered 
originally by Fukugita and Yanagida and self-energy 
diagrams
as first introduced and discussed in Ref.~\cite{fps}. 
%%%%%%%%%%%%%%%%%%%%%%%%%%%%%%%%%%%%%%%%%%%%%
%
\begin{figure}[t]
\begin{center}
\begin{picture}(270,60)(0,0)
%1st diagram
\Line(0,30)(30,30)
\DashArrowLine(30,30)(60,60){5}
\ArrowLine(30,30)(60,0)
\Text(10,22)[]{$N_k$}
\Text(47,4)[]{$l_j $}
\Text(47,55)[]{$\phi^\ast $}
%2nd diagram
\Line(70,30)(100,30)
\DashArrowLine(130,60)(160,60){5}
\ArrowLine(130,0)(160,0)
\Line(130,60)(130,0)
\ArrowLine(130,60)(100,30)
\DashArrowLine(100,30)(130,0){5}
\Text(75,22)[]{$N_k$}
\Text(105,12)[]{$\phi$}
\Text(107,51)[]{$ l^c_i$}
\Text(137,30)[]{$N_l$}
\Text(155,52)[]{$\phi^\ast$}
\Text(155,8)[]{$ l_{j}$}
%3rd diagram
\Line(170,30)(190,30)
\DashArrowArc(205,30)(15,180,360){5}
\ArrowArc(205,30)(15,0,180)
\Line(220,30)(240,30)
\DashArrowLine(240,30)(270,60){5}
\ArrowLine(240,30)(270,0)
\Text(175,22)[]{$N_k$}
\Text(205,6)[]{$\phi$}
\Text(206,55)[]{$ l^c_i$}
\Text(230,22)[]{$N_l$}
\Text(257,55)[]{$\phi^\ast$}
\Text(257,4)[]{$ l_{j}$}
\end{picture}
\end{center}
\caption{Tree level and loop diagrams interfering together.}
\label{fig}
\end{figure}
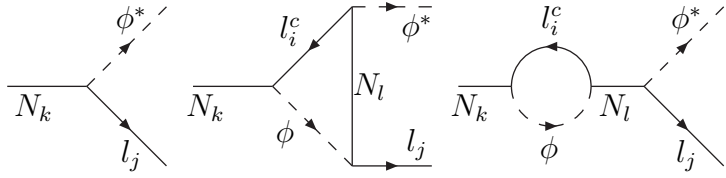
%%%%%%%%%%%%%%%%%%%%%%%%%%%%%%%%%%%%%%%%%%%%%
The lepton asymmetry induced by these diagrams,
\begin{equation}
\varepsilon_{N_{k}}=\frac{\Gamma(N_{k} \rightarrow L_j + \Phi^\dagger)
-\Gamma(N_{k} \rightarrow L^\dagger_j + \Phi)}{\Gamma_{N_{k}}} \,,
\label{epsfy}
\end{equation}
which is nothing but 
the averaged amount of lepton number which is created per 
decay of right-handed neutrinos,
is then obtained as:
\begin{eqnarray}
\varepsilon^V_{N_k} &=& - \frac{1}{8 \pi} \sum_{l}
\frac{\sum_{i,j}  \mbox{Im}[
h_{ki}^{\ast} h_{kj}^{\ast}
h_{li} h_{lj}
]}
{ \sum_{i} |h_{ki}|^2  }
\
\sqrt{x_l} \Big[  (1+x_l) \mbox{Log} (1+1/x_l) -1 \Big] ,
\label{epsVfy} \\
\varepsilon^S_{N_k} &=& - \frac{1}{8 \pi} \sum_{l}
\frac{\sum_{i,j}  \mbox{Im}[
h_{ki}^{\ast}h_{kj}^{\ast}
h_{li} h_{lj}
 ]}
{ \sum_{i} |h_{ki}|^2  }
\sqrt{x_l} (x_l -1)^{-1} \, ,
\label{epsSfy}
\end{eqnarray}
where 
$\epsilon^V_{N_k}$ and 
$\epsilon^S_{N_k}$
are the contribution of the vertex and self-energy diagrams respectively, with
$x_l=(M_{N_l} / M_{N_k})^2$. This leads to a ratio 
$n_L/s 
\simeq \sum_k (\varepsilon^V_{N_k} 
+ \varepsilon^S_{N_k}) n_\gamma/2s  
\sim \sum_k (\varepsilon^V_{N_k} 
+ \varepsilon^S_{N_k})/ g_\ast$ with $n_\gamma$ the photon number density and
$g_\ast$ the number of active degrees of freedom at these temperatures.
The sphalerons which are active at temperatures between $\sim 100-200$~GeV 
and $\sim 10^{12}$~GeV will convert approximately one third 
of this lepton asymmetry
to a baryon asymmetry. $n_B/s$ is given by \cite{ht}: 
\begin{equation}
\Big(\frac{n_B}{s}\Big)_{fin} = \Big( \frac{8 N_F + 4 N_H}{22 
N_F + 13 N_H} \Big) 
\Big(\frac{n_B-n_L}{s}\Big)_{init} \, ,
\label{hatu}
\end{equation}
where $N_H$ is the number of Higgs doublets and $N_F$ the number of 
fermionic families.
However, in order for this lepton asymmetry to be 
effectively created in the thermal evolution 
of the universe, the decays must take place 
out-of-equilibrium. 
This will be the case if the decay rate is smaller than 
the expansion rate of the universe, parametrized in terms of the Hubble 
constant $H$ \cite{za}:
\begin{equation}
\Gamma_{N_{k}} < H(T=M_{N_{k}})=
\sqrt{\frac{4 \pi^3 g_\ast}{45}} \frac{T^2}{M_{\footnotesize{Planck}}} 
\Big|_{T=M_{N_k}} \, .\label{ooecfy}
\end{equation}
Since this inequality is mediated by the Planck 
scale $M_{Planck} \sim 10^{19}$~GeV, it is
easier to satisfy it at scales not many orders 
of magnitude smaller than this scale.
To satisfy it at even smaller scales, 
values of the Yukawa couplings much smaller 
than unity will have to be considered.

If Eq.~(\ref{ooecfy}) is not satisfied or 
in presence of scattering effects which may 
damp the asymmetry, an explicit calculation of the evolution of the asymmetry 
from the Boltzmann equations is required. Since for most of the
realistic models of leptogenesis at the 1-10 TeV scale, discussed below,
such a detailed calculation is necessary, we explicitly 
outline the calculation
in appendix A.

\newpage
%
%%%%%%%%%%%%%%%%%%%%%%%%%%%%%%%%%%%%%%%%%%%%%%%%%%%%%%%%%%%%%%%%%%%%%%%
%
\section{Why it is difficult to construct a simple leptogenesis 
model at the TeV scale \label{frame}}
Trying to build a leptogenesis model at the 1-10 TeV scale
the main difficulties we have to face are the following:
\begin{itemize}
\item First the out-of-equilibrium condition for the 
decay width, Eq.~(\ref{ooecfy}),
imposes the general condition that the couplings
are very tiny. This is due to the facts that first, as was just mentioned, 
this condition is mediated by the 
Planck scale and secondly, the decay width is in general only linear 
in the mass of 
the decaying particle in contrast to the Hubble constant
which depends quadratically on
this mass.
This means that the product of couplings entering the decay width has to be 
as much as 10 orders of magnitude smaller at the TeV scale than at 
the $10^{13}$ GeV scale. 
Beyond the fact that the naturality of such tiny couplings 
can be questionable, the 
major problem is that it generally induces a very tiny asymmetry 
due to the fact that the asymmetry 
in most possible models is proportional to the same tiny couplings. 
For example 
in the Fukugita-Yanagida model, 
to naturally satisfy
Eq.~(\ref{ooecfy}) together with getting 
$\varepsilon_{N}/g_\ast$ of order $10^{-10}$, we need typically 
$M_{N}$ above $10^{10}$~GeV.
With smaller value of $M_{N}$, 
for example with $M_N \sim 10$~TeV, the condition 
of Eq.~(\ref{ooecfy}) imposes that the Yukawa couplings have 
to be smaller than $\sim 10^{-6}$. Inserting this value of the couplings in
Eqs.~(\ref{epsVfy})-(\ref{epsSfy}) we 
obtain from the couplings alone a factor of $\sim 10^{-12}$
in the asymmetry, which makes the asymmetry at least $\sim 6$ orders of 
magnitude too small. By taking 
larger couplings we could obtain a larger asymmetry $\varepsilon_N$
but the final value of $n_L/s$ will 
be damped by an extra factor (which is larger 
than the factor gained in $\varepsilon_N$) from the inverse decay processes.
Consequently we don't gain anything.
Possible mechanisms to remedy this problem are discussed in section 4.

\item At the TeV scale, various 
scatterings can also be very fast with respect to the 
Hubble constant. 
The scatterings 
directly proportional to the couplings which are 
constrained by the out-of-equilibrium condition on the decay
have in general a relatively small damping effect.
However, other scatterings can be 
very fast and damp largely the asymmetry. This is particularly the case 
with gauge scatterings if the particles producing the asymmetry 
are not neutral or $SU(2)_L$ singlets. To illustrate this fact let us 
take the example of a charged $SU(2)_L$ singlet scalar $S^+$ whose 
decay would be at the origin of a large asymmetry.
The lepton number conserving $S^+ S^- \leftrightarrow \gamma \gamma$ scattering 
is a very fast process 
which would damp the asymmetry by several orders of magnitude. 
Integrating the Boltzmann equation (Eq.~(\ref{bol1}) in appendix A), with a mass
$m_S\sim 1$~TeV, we observe that,
due to this scattering, there is no substantial departure from equilibrium down 
to $T\sim 50$~GeV.\footnote{As well known this 
freeze-out temperature can also be obtained
by looking for the temperature at which the scattering rate 
$\Gamma_{scatt}=\gamma_{scatt}^{eq}/n_S^{eq}$ is of 
order the Hubble constant [see Eqs.~(\ref{maxbol}) and (\ref{gamscatt})].}
At this freeze-out temperature, the ratio of the number 
density of $S$ particles
to the entropy (given by 
Eqs.~(\ref{entropy}) and~(\ref{maxbol}) in appendix A) is strongly 
Boltzmann suppressed: we 
get $n_S^{eq}/s \sim 3 \cdot 10^{-10}$
to be compared with $n_S^{eq}/s \sim 2 \cdot 10^{-3}$ 
at $T=m_S \sim 1~\hbox{TeV}$.
We therefore expect that the asymmetry 
is suppressed by about six-seven orders of magnitude!\footnote{Actually 
this calculation gives an order of magnitude 
estimate but the exact suppression factor
depends on the value of $K_S=\Gamma_S/H(T=m_S)$. For large value of $K_S$ (i.e.
$K_S >>1$),
some scalar singlets will have the time to decay before thermalizing well
above $T \sim 50$~GeV, so that the suppression factor will be less important.
However in this case
the inverse decay processes will damp
largely the asymmetry so that we also end up with a several orders
of magnitude suppression of the lepton asymmetry.} To
compensate this suppression we would need a large enhancement
in addition to the one already required to solve the problem
of the previous paragraph.
We conclude from this fact that, at such low scales, 
the particle at the origin of the 
asymmetry be better neutral and a gauge singlet of any 
low-energy non-abelian gauge symmetry.
In this respect a right-handed neutrino (with heavier right-handed W if 
this model is embedded in a left-right model) is a particularly suitable 
candidate.
A neutral gauge singlet scalar would also be a candidate although it is 
in general not so easy to introduce a lepton number violation with such a 
candidate.\footnote{Except at least with right-handed sneutrinos in 
supersymmetric models. 
For a leptogenesis model based on sneutrino decay at higher temperatures see
Ref.~\cite{mur}.}

\item In the more ambitious and more interesting case where the source 
of lepton number violation at the origin of the asymmetry is also at the 
origin of the neutrino oscillations, an other problem could in
general occur. Two cases have to be distinguished:
\begin{itemize}
\item First in the case where the 
neutrino masses are produced at tree level, as 
in the seesaw mechanism with right-handed neutrinos, 
the values
of the couplings which are needed to generate neutrino masses are generally
slightly larger than 
the ones allowed by the out-of-equilibrium condition.
For example, with right-handed 
neutrinos with masses of order 10 TeV the typical 
value of the $h_{ij}$ Yukawa 
couplings which is needed to have a neutrino mass of order $10^{-1}$~eV
is $\sim 5 \cdot10^{-6}$, that is to say 
about 1 order of magnitude larger than the bound from 
the out-of-equilibrium condition on the decay 
width of Eq.~(\ref{ooecfy}). This 
will induce a relatively large damping 
effect from inverse decays [$K_N$ will be of order 
100 in Eq.~(\ref{bol1})-(\ref{bol2})]. 
This effect could be possibly compensated by an enhancement 
mechanism if any but this 
is in general not possible (except possibly with the 
"mass degeneracy" mechanism introduced in section 4 below). 
%However 
%if this neutrino mass is not generated by the lightest right-handed 
%neutrino, whose decay is at the origin of 
%leptogenesis, but is generated by the two heavier
%right-handed neutrinos, the neutrino constraints are in general no problem for
%leptogenesis.

\item Secondly in the case of neutrino masses generated by radiative 
processes, as in the
Zee model or in R-parity violating supersymmetric 
models, it is quite difficult
to generate the neutrino masses and the lepton asymmetry from the same 
interactions. The 
reason is that to generate in those models a neutrino mass of at least
the SuperKamiokande bound we need couplings several orders of
magnitude larger than the upper bound on these couplings from the 
out-of-equilibrium condition.
For example the trilinear R-parity violating 
couplings
needed for neutrino 
masses are typically of order $10^{-4}$ \cite{hempf} while the 
out-of-equilibrium condition
on the associated L-violating 2 body scatterings requires
couplings of order $10^{-7}$ \cite{erase,rms}. 
The out-of-equilibrium condition for these scatterings is then violated 
by 6 orders of magnitude which induces a huge suppression of the
associated asymmetry,
which would be very difficult if not impossible 
to compensate 
by any enhancement effect \cite{rms}.\footnote{In the case where the 
neutrino masses are not generated by the trilinear R-parity violating terms
but by the bilinear R-parity violating terms 
the discussion is different but, as 
explained in Ref.~\cite{hms2}, this 
leads to a far too small lepton asymmetry.} There are 
nevertheless some ways for avoiding those 
suppressions due to neutrino constraints 
as will be explained below.
\end{itemize} 
\end{itemize} 

%
%%%%%%%%%%%%%%%%%%%%%%%%%%%%%%%%%%%%%%%%%%%%%%%%%%%%%%%%%%%%%%%%%%%%%%%
%
\section{Three possible enhancement mechanisms 
\label{ana}}
To satisfy approximately the out-of-equilibrium condition on the decay width
together with 
inducing a 
large enough asymmetry we will consider three possible enhancement mechanisms:
\begin{enumerate}
\item {\it Mass degeneracy}: It has been observed in Ref.~\cite{fps} that 
asymmetries induced by self-energy diagrams (as the 
third diagram of Fig.~1) display an 
interesting resonant behavior when the masses of the decaying particles are
nearly degenerate. In the Fukugita-Yanagida model this resonant behavior
occurs when at least 2 right-handed neutrinos are nearly 
degenerate [see Eq.~(\ref{epsSfy})].
This resonance effect
turns out to be maximum for a mass difference of order the decay 
width.\footnote{A more careful calculation of the asymmetry based 
on a resummed calculation gives approximately the same result \cite{pil}.} 
In this way, starting with right-handed neutrinos with 
masses of order $\sim 10$ TeV, and with Yukawa 
couplings of order $10^{-6}$, i.e. 
satisfying approximately the out-of-equilibrium condition 
of Eq.~(\ref{ooecfy}), one can 
in principle get a many orders of magnitude enhancement by requiring 
enough degeneracy. The degree of 
degeneracy which is required at this scale, if smaller than the 
one corresponding to the resonance condition, is 
nevertheless huge \cite{pil}: $(m_{Ni}-m_{Nj})/(m_{Ni}+m_{Nj}) \sim 10^{-7}$!
The naturality of such a degeneracy is highly debatable. Note 
also that the 
perturbativity of such a huge enhancement has been questioned 
in Ref.~\cite{bp}.

\item {\it Hierarchy of couplings}: Another possibility
assumes 
two particles decaying to the same decay products. The lighter one (which 
we denote by "$A$") and the heavier one ("$B$") couple
to these decay products with 
couplings "$g_A$" and "$g_B$" respectively. The $g_A$ couplings 
are taken very suppressed in order that $A$ decays 
out-of-equilibrium at the 1-10 TeV scale. It is this decay
which is 
at the origin of the asymmetry. The couplings $g_B$ of the heavier 
particle
are on the other hand taken unsuppressed (and are eventually 
large enough to be at the origin of the neutrino masses). At temperature of 
order $m_A$ we can take $m_B$ large enough for all particles
$B$ to have decayed away (a factor $m_B/m_A$ of $\sim 3-10$ is in general 
enough because 
the number of particle "B" is fastly Boltzmann suppressed at temperature 
below its mass). 
With these assumptions a large asymmetry can be produced from the fact that 
the one loop diagrams similar to the ones of Fig.~1 for the decay of A 
with a virtual 
B will give an asymmetry proportional to $(g_A g_B)^2/g_A^2=g_B^2$
which is unsuppressed. This simple and attractive 
mechanism has been discussed in the framework 
of a R-parity violating model in Ref.~\cite{hms1,hms2}. The 
naturality of the hierarchy of
couplings between two particles $A$ and $B$ which are very similar
since they couple to the same decay products is nevertheless debatable. The 
typical value of the ratio $g_A/g_B$ which is needed is of 
order $10^{-3}$ \cite{hms2}.

\item {\it Three body decays}: Three body decays appear to 
be very interesting and more natural for
generating a lepton asymmetry at the 1-10 TeV scale. First because, to satisfy
the out-of-equilibrium condition on the decay width, three body decays
are naturally smaller than two body decays due to phase space suppressions 
and the fact that they naturally involve more couplings. Secondly because, 
to induce a large enough asymmetry, they don't
require any special hierarchy of couplings or mass degeneracy. 
The mechanism is the following: 
let us assume a trilinear coupling $g_1$ between three particles $A$, $B$, $C$ 
with $m_B<m_A<m_C$ and let us assume that $C$ in addition to its decay to $A+B$ 
can also decay to lighter particles $D+E$ through some couplings $g_2$. 
Then $A$ can decay only to the
three body decay $A \rightarrow B+C^\ast \rightarrow B+D+E$ 
through a virtual C. 
In this context it is easy to see that a large asymmetry can be produced.
The point is that the out-of-equilibrium condition 
on the decay width will give an upper bound on a quartic expression
in the couplings (i.e. on $g_1^2 g_2^2$) but the
asymmetry will be in general only quadratic in these couplings. This has to 
be compared with the usual 
two body decays where both the asymmetry 
and the decay width are quadratic in the couplings.
For example if the asymmetry comes from loop diagrams not
involving the $A-B-C$ couplings $g_1$ but various $C-D-E$ 
couplings $g_2$
the asymmetry
will be proportional to $g_1^2 g_2^4/(g_1^2 g_2^2)\sim g^2_2$ and can be 
very large even if $g_1^2 g_2^2$ has to be small to satisfy the 
out-of-equilibrium decay condition. Note that no
special hierarchy of couplings has to be assumed. For example, taking 
all couplings $g_1$ and $g_2$ of order $10^{-3}$-$10^{-4}$ the asymmetry
can be naturally large enough ($g_2^2\sim 10^{-6}$-$10^{-8}$) 
with a sufficiently suppressed 
decay ($g_1^2 g_2^2 \sim 10^{-12}$-$10^{-16}$). Therefore, if as explained above
the natural scale for producing a large 
asymmetry 
with the usual
two body decays 
is around $\sim 10^{10}$~GeV or above,
with three body decays the 1-10~TeV scale is a perfectly natural scale
for producing such a large asymmetry!
No special "trick" has to be used to enhance the asymmetry at low scale as 
with the usual two body decays and the two other mechanisms.
Note in addition that, if a hierarchy of couplings is in general 
not necessary, an even much larger 
asymmetry can be obtained by taking the 
hierarchy $g_2 > g_1$.\footnote{In other words,
if the lepton number violation lies in the $g_2$ couplings this 
means that the amount of lepton number violation is not constrained to be 
small anymore by the out-of-equilibrium condition.}
Moreover as we will see with an 
example in the next section the values of the $g_1$ and/or $g_2$ couplings 
which 
are required for leptogenesis have typically the size required 
to generate radiatively neutrino masses with the right orders of magnitude.
Note however that in this framework it must be still checked that no
two body scatterings, which can be naturally large with such values of the 
couplings, can erase the asymmetry. In the next section we will see with an 
explicit example that this problem can be in general avoided easily.
The point is that, with this mechanism, even if at $T \sim m_C$ there 
exist in general very stringent bounds on the coupling $g_2$ from 
imposing that associated scatterings don't erase any lepton 
asymmetry \cite{erase,fy1,dr}, those
bounds can be considerably relaxed. The asymmetry produced by the 3 body 
decay of particle $A$ will not be erased by those scatterings even 
for much larger values of $g_2$ because it is produced 
at $T \sim m_A$ which is below $m_C$ (i.e. when the $C$ particles have already
decayed away).

\end{enumerate}

We close this section by briefly reviewing the possibility to implement these 
mechanisms in the framework of well-known neutrino mass models such as the 
Fukugita-Yanagida model, the triplet Higgs model \cite{triplet,hms3}, 
the Zee model [32-36]
%\cite{zee,otherzee,kan,grim,framp} 
and the 
models with R-parity violation \cite{hempf}.
\begin{itemize}
\item {\it The Fukugita-Yanagida model}: here, as said 
above, the mass degeneracy 
mechanism has been extensively discussed in 
the literature [10-12].
%\cite{fps,pil,bp} 
On the other hand the three body decay mechanism and the hierarchy 
of coupling mechanism don't appear to be very helpful here.
The former mechanism would 
require the existence of heavier particle which are not 
present in this model. The 
later mechanism
would require a neutrino $N_1$ with mass $M_1$ and suppressed 
Yukawa couplings $h_1$ which is lighter than 
another one $N_2$ with mass $M_2$ and unsuppressed 
Yukawa couplings $h_2$. In this case, 
from Eqs.~(\ref{epsVfy})-(\ref{epsSfy}),  $\varepsilon_{N_1}$
would be proportional 
to $(1/8 \pi) h_2^2 M_1/M_2$ (neglecting higher order terms in $(M_1/M_2)^2$). 
However, 
assuming that the neutrino 
masses cannot be much larger than $\sim 1$~eV,
the combination $h_2^2v^2/M_2$ cannot be much larger than 1~eV.
This gives an upper bound on $\varepsilon_{N_1}$ which is given
by:
\begin{equation}
\varepsilon_{N_1}  \lsim \frac{1}{4\pi} \frac{m_\nu M_{N_1}}{v^2}
\end{equation}
and which for example with 
$M_{1} \sim 10$ TeV,
is at most of order $10^{-11}$, that is to say anyway at least 3 orders 
of magnitude too small. To obtain a value of $\varepsilon_{N_1}$ at 
least of order $\sim 10^{-8}$, which is necessary 
to have $n_L/s \sim 10^{-10}$, 
we need a value of $M_{N_1}$ above the bound:
\begin{equation}
 M_{N_1} \gsim 4 \pi v^2\varepsilon_{N_1}/m_\nu \sim 10^7 \hbox{GeV}. 
\end{equation}
\item {\it The Higgs triplet model}: in this model the neutrino masses 
are also generated through a seesaw mechanism. Therefore 
the hierarchy of couplings 
mechanism is not helpful in the same way as for
the Fukugita-Yanagida model. The three body decay mechanism is also not possible
because here too it would require additional particles which are not present 
in this model. 
The mass degeneracy mechanism on the other hand could be useful 
for the triplets 
as for the right-handed neutrinos.
However, in contrast with the right-handed neutrinos, the Higgs triplets are 
not $SU(2)_L$ singlets. Therefore, as discussed 
in Ref.~\cite{hms3}, if the leptogenesis mechanism 
is quite natural for triplet masses
above $10^{9}-10^{10}$ GeV, for lower masses the damping effect of
the associated gauge scatterings (i.e. of a triplet 
pair to a gauge boson pair) becomes important and
it appears to be impossible to have a successful mechanism at scales as low 
as 1-10 TeV. 
The degree of degeneracy required would be larger than the 
resonance condition would allow.\footnote{Of course 
a low value of the triplet mass would be welcome to avoid 
large corrections to the mass of the Higgs 
doublet coming from their self-energy
with an internal heavy triplet. 
This is however not possible as just explained.}

\item {\it The Zee model}: in the Zee model the three body decay mechanism 
cannot be used  because here too it requires additional particles 
which are not present in this model.
To avoid 
the problems related to the neutrino masses explained in section 3,
the hierarchy of couplings mechanism could be used.\footnote{This 
would require 
the introduction of a second charged scalar singlet
in order to have CP-violation, both scalar singlets playing the role
of particle "A" and "B" in the hierarchy of coupling mechanism explained 
in section 3. 
The one loop (self-energy) diagram responsible for 
the asymmetry would be the same as in the 
triplet model replacing triplets by singlets.} However
similar to the triplet case the gauge scatterings will considerably damp 
the asymmetry. In the Zee model the main effect comes from the
$S^+ S^- \leftrightarrow \gamma \gamma$ scatterings discussed previously
in section~3. 

\item {\it The supersymmetric models with R-parity violation}: in these 
models the neutrino masses can be induced by the R-parity and
L violating terms. To induce the leptogenesis from the same terms 
we could use the hierarchy of 
couplings mechanism or the three body decay mechanism.
By using them we could avoid the problems related to the neutrino 
masses explained in section 3.
However here too the gauge scatterings would damp the asymmetry largely by
rendering the decaying particles
in thermal equilibrium down to temperature
far below their mass.
For example the decays of the sfermions or of the charginos couldn't 
produce a large enough asymmetry due to the effect of the scatterings of
a sfermion pair or of a chargino pair to a gauge boson pair.
For a neutralino, if this neutralino 
is essentially a wino $\tilde{W}_3$ or a Higgsino, 
the scattering of 2 neutralinos going to 2 $W$
mediated by a chargino will have a very large damping 
effect in the same way.\footnote{We thank S. Davidson for pointing us the 
potential effect of this scattering.
Calculating explicitly its effect, which was not taken into account
in Ref.~\cite{as2}, it can be checked that it changes drastically 
the results of this reference. It also turns out that this effect
was underestimated in Ref.~\cite{hms1,hms2} for which a reanalysis
of the parameter space in this model should be performed.}
A similar comment
applies to the case of a bino (as well as to the case of a wino) 
through the squark mediated 
scatterings of 2 binos 
to a quark pair.
\end{itemize}

In summary except for the debatable mass degeneracy mechanism in the
Fukugita-Yanagida model, there is no simple leptogenesis 
solutions at low scale in all these models.

\newpage
%%%%%%%%%%%%%%%%%%%%%%%%%%%%%%%%%%%%%%%%%%%%%%%%%%%%%%%%%%%%%%%%%%%%%%%%%%
\section{An explicit model based on three body decays}

In the following, based on the most natural 3 body decay mechanism, 
we build an alternative and successful model 
which avoids the problems of the existing neutrino mass models just explained.

%%%%%%%%%%%%%%%%%%%%%%%%%%%%%%%%%%%%%%%%%%%%%%%%%%%%%%%%%%%%%%%%%%%%%
\subsection{The model}
%%%%%%%%%%%%%%%%%%%%%%%%%%%%%%%%%%%%%%%%%%%%%%%%%%%

Probably the 3 body decay enhancement mechanism explained in
section 4
could be implemented in many different contexts.\footnote{See
also Ref.~\cite{as2} for a non-leptogenesis baryogenesis model
based on the R-parity violating 3 body decays of a neutralino and
Refs.~\cite{mr,as} for leptogenesis models at the electroweak phase 
transition based also on R-parity violating 3 body decays of neutralinos.}
However as 
explained above to avoid large damping effects of gauge scatterings, the decay 
should be from a neutral gauge singlet, which restricts the possibilities.
The most natural candidate is the right-handed neutrino which we will consider.
To implement the 3 body decay mechanism, a heavier particle has to be 
introduced.
One simple and minimal 
possibility is to assume additional charged scalar singlets.
In the following we will assume two scalar singlets $S_{1,2}^+$. We
will also assume
two Higgs doublets
$H_k\equiv ( \phi^0_k \,\, \phi^-_k)^T$, $k=1,2$ as in the 
Zee model~[32-36].
% \cite{zee,otherzee,kan,grim,framp}.
In this context, beside the usual Yukawa couplings
of the right-handed neutrinos of Eq.~(\ref{Lfy}) 
and beside the Zee couplings of these singlets to two leptons or 
to two scalar doublets,
the right-handed neutrinos can couple to
a charged scalar singlet and a right-handed lepton. In a general way
we have therefore the following interactions:
\begin{equation}
{\cal L}_{Y} \owns
 h^k_{Lij} \bar{L}_j \psi_{Ri} H_k  
 + h^k_{Rij} l^T_{Rj} C^{-1} \psi_{Ri} S^+_k  
+ f^k_{ij} L_i^T C^{-1} i \tau_2 L_j S^+_k
+ \lambda^k_S H_1^T i \tau_2 H_2 S_k^+ + h.c.
\label{L3bd}
\end{equation}
where 
$L_i \equiv ( \nu_i \,\, l_{Li}^-)^T$.
From this it is easy to build a successful model of leptogenesis.

%%%%%%%%%%%%%%%%%%%%%%%%%%%%%%%%%%%%%%%%%%%%%%%%%%%%%%%%%%%%%%%%%%
\subsection{Leptogenesis}
%%%%%%%%%%%%%%%%%%%%%%%%%%%%%%%%%%

For leptogenesis we assume that 
the right-handed neutrinos decay to a Higgs doublet 
and a left-handed lepton with suppressed couplings $h^k_{Lij}$ 
in order to satisfy 
approximately the out-of-equilibrium condition on the associated 
two body decay widths. Assuming right-handed neutrinos at low scales 
these couplings have anyway to be tiny not to 
induce too large masses for the light neutrinos.
The right-handed neutrinos 
can also decay to a right-handed anti-lepton 
plus 2 left-handed leptons or 2 scalar doublets via a scalar singlet as shown
in Fig.~2.
\begin{figure}[t]
\begin{center}
\begin{picture}(190,84)(0,0)
%1st diagram
\Line(0,50)(33,50)
\ArrowLine(63,80)(33,50)
\DashArrowLine(33,50)(53,30){5}
\ArrowLine(53,30)(81,25)
\ArrowLine(53,30)(58,2)
\Text(5,42)[]{$N_i$}
\Text(63,68)[]{$l^c_j $}
\Text(52,46)[]{$S^-_{k} $}
\Text(81,32)[]{$l_m $}
\Text(65,3)[]{$l_n $}
%2nd diagram
\Line(105,50)(138,50)
\ArrowLine(168,80)(138,50)
\DashArrowLine(138,50)(158,30){5}
\DashArrowLine(158,30)(186,25){5}
\DashArrowLine(158,30)(163,2){5}
\Text(110,42)[]{$N_i$}
\Text(168,68)[]{$l^c_j $}
\Text(157,46)[]{$S^-_{k} $}
\Text(187,33)[]{$\phi_{1,2}^0 $}
\Text(174,4)[]{$\phi_{2,1}^- $}
\end{picture}
\end{center}
\caption{Right-handed neutrino three body decays.}
\label{fig2}
\end{figure}
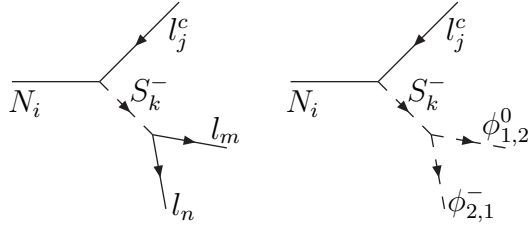
We assume that these three body decay partial widths also 
satisfy the out-of-equilibrium condition 
due to three body decay suppression and the fact that we assume
that the couplings $h^k_{Rij}$, $f_{ij}^k$ and $\lambda_S^k$
are small enough (without necessarily being tiny). 
In this way the asymmetry is obtained naturally large
as explained in section 4. 
To see that let us first write down the $N_i$ decay widths:
\begin{eqnarray}
\Gamma_{N_{i}} &\simeq& \frac{1}{8 \pi} \sum_{jk} |h^k_{Lij}|^2 M_{N_{i}} 
\nonumber\\
&&+
\frac{1}{(2 \pi)^3}\frac{1}{48} 
\sum_{jkl} h^k_{Rij} h^{l\ast}_{Rij}
\frac{M_{N_i}^3 \lambda_S^{k\ast} \lambda_S^l}
{m^2_{S_k} m^2_{S_l}} \nonumber \\
&&+\frac{1}{(2 \pi)^3} \frac{1}{96} \sum_{jklmn} h^k_{Rij} h^{l\ast}_{Rij}
f^{k\ast}_{mn} f^l_{mn} \frac{M_{N_i}^5}{m^2_{S_k} m^2_{S_l}} \, .
\label{gam3bd}
\end{eqnarray}
In this equation we have neglected all light fermions and scalar masses 
and we kept only the highest term in $M_{N_i}^2/m_{S_k}^2$ (k=1,2; i=1,2,3).
The asymmetry induced by the three body decays is given by the 
diagrams of Fig.~2. Calculating the 
interference of tree level and one-loop diagrams we get:
\begin{eqnarray}
\varepsilon_{N_i}&\simeq& A_{N_i}
\sum_{j,m,n}
\Big[\,
\hbox{Im}[h^{2 \ast}_{Rij} h^{1}_{Rij} 
\lambda_S^{1 \ast} \lambda_S^{2} ]
\Big(
\frac{|f^1_{mn}|^2}{m^2_{S1}}-\frac{|f_{mn}^2|^2}{m^2_{S2}} \Big) \, 
\nonumber \\
&&\,\,\,\, + \, \hbox{Im}[ h^2_{Rij} 
h^{1 \ast}_{Rij} f_{mn}^1 f_{mn}^{2 \ast} ] 
\Big( 
\frac{|\lambda_S^1|^2}{m_{S1}^2} - \frac{|\lambda_S^2|^2}{m^2_{S2}} \Big)
\nonumber \\
&&\,\,\,\, + \,  \hbox{Im}[f_{mn}^{2} f_{mn}^{1 \ast} 
\lambda_S^1 \lambda_S^{2 \ast} ]
\Big(\frac{|h_{Rij}^1|^2}{m^2_{S_1}} - \frac{|h_{Rij}^2|^2}{m^2_{S_2}} 
\Big) \Big] \, ,
\label{eps3bd}
\end{eqnarray}
with:
\begin{equation}
A_{N_i}=\frac{1}{\Gamma_{N_i}}
\frac{1}{(2 \pi)^3 } \frac{1}{12} \frac{\pi}{(4 \pi)^2}
\frac{M_{N_i}^5}{m^2_{S_1} m^2_{S_2}} \,.
\label{eps3bdA}
\end{equation}
In this asymmetry there are several 
combinations of couplings which provide the CP-violating phases.
%%%%%%%%%%%%%%%%%%%%%%%%%%%%%%%%%%%%%%%%%%%%%
%
%
\begin{figure}[t]
\begin{center}
\begin{picture}(258,124)(0,0)
%1st diagram
\Line(0,90)(33,90)
\ArrowLine(63,120)(33,90)
\DashArrowLine(33,90)(53,70){5}
\DashArrowArc(63,60)(14,135,325){5}
\DashArrowArcn(63,60)(14,135,325){5}
\DashArrowLine(73,50)(93,30){5}
\ArrowLine(93,30)(121,25)
\ArrowLine(93,30)(98,2)
\Text(5,82)[]{$N_i$}
\Text(63,108)[]{$l^c_j $}
\Text(38,72)[]{$S^-_{k} $}
\Text(82,78)[]{$\phi_{1,2}^0 $}
\Text(47,43)[]{$\phi_{2,1}^- $}
\Text(78,32)[]{$S^-_{l} $}
\Text(119,33)[]{$l_m $}
\Text(105,3)[]{$l_n $}
%2nd diagram
\Line(135,90)(168,90)
\ArrowLine(198,120)(168,90)
\DashArrowLine(168,90)(188,70){5}
\ArrowArc(198,60)(14,135,325)
\ArrowArcn(198,60)(14,135,325)
\DashArrowLine(208,50)(228,30){5}
\DashArrowLine(228,30)(256,25){5}
\DashArrowLine(228,30)(233,2){5}
\Text(140,82)[]{$N_i$}
\Text(198,108)[]{$l^c_j $}
\Text(173,72)[]{$S^-_{k} $}
\Text(214,77)[]{$l_m $}
\Text(185,44)[]{$l_n $}
\Text(213,32)[]{$S^-_{l} $}
\Text(256,34)[]{$\phi_{1,2}^0 $}
\Text(245,3)[]{$\phi_{2,1}^- $}
\end{picture}
\end{center}
\caption{Loop diagrams interfering with the  $N_i$ tree decay.}
\label{fig3}
\end{figure}
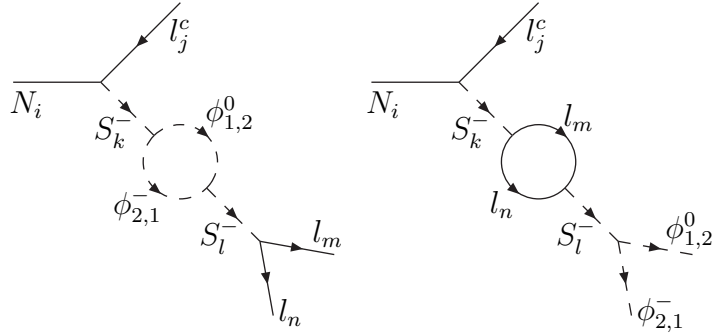
Those phases cannot be absorbed in a redefinition of couplings.
Note that with only one scalar singlet these phases would vanish. We
need therefore at least two scalar singlets. Similarly 
we need two scalar doublets
because with only one we wouldn't have any couplings of the scalar singlets
with two scalar doublets,
hence no asymmetry.

Actually in this model the three body decay asymmetry of Eq.~(\ref{eps3bd})
doesn't give the full result. We have also to take into account the effect
of the small mixings 
between the light charged scalars $\phi_{1,2}^{\pm}$ and the heavy
charged singlets induced by the vacuum expectation values of 
the neutral components
of both Higgs doublets. Due to these vevs and associated mixings  
the three body decays of Fig.~3 lead also
to the $N_i \rightarrow l_j^c + \phi_{1,2}^-$ two body decays 
as shown in Fig.~4.
%%%%%%%%%%%%%%%%%%%%%%%%%%%%%%%%%%%%%%%%%%%%%
%
%
\begin{figure}[t]
\begin{center}
\begin{picture}(258,124)(0,0)
%1st diagram
\Line(0,70)(33,70)
\ArrowLine(63,100)(33,70)
\DashArrowLine(33,70)(53,50){5}
\DashArrowLine(53,50)(73,30){5}
\DashLine(53,50)(65,62){4}
\Line(61,62)(69,62)
\Line(65,58)(65,66)
\Text(5,62)[]{$N_i$}
\Text(63,88)[]{$l^c_j $}
\Text(38,52)[]{$S^-_{k} $}
\Text(81,70)[]{$\langle \phi_{1,2}^0 \rangle $}
\Text(84,35)[]{$\phi_{2,1}^- $}
%2nd diagram
\Line(135,90)(168,90)
\ArrowLine(198,120)(168,90)
\DashArrowLine(168,90)(188,70){5}
\ArrowArc(198,60)(14,135,325)
\ArrowArcn(198,60)(14,135,325)
\DashArrowLine(208,50)(228,30){5}
\DashArrowLine(228,30)(248,10){5}
\DashLine(228,30)(240,42){4}
\Line(236,42)(244,42)
\Line(240,38)(240,46)
\Text(140,82)[]{$N_i$}
\Text(198,108)[]{$l^c_j $}
\Text(173,72)[]{$S^-_{k} $}
\Text(214,77)[]{$l_m $}
\Text(186,42)[]{$l_n $}
\Text(213,32)[]{$S^-_{l} $}
\Text(256,50)[]{$\langle \phi_{1,2}^0 \rangle $}
\Text(259,15)[]{$\phi_{2,1}^- $}
\end{picture}
\end{center}
\caption{Two body decays induced by the Higgs vevs and loop diagram
contributing to the asymmetry.}
\label{fig4}
\end{figure}
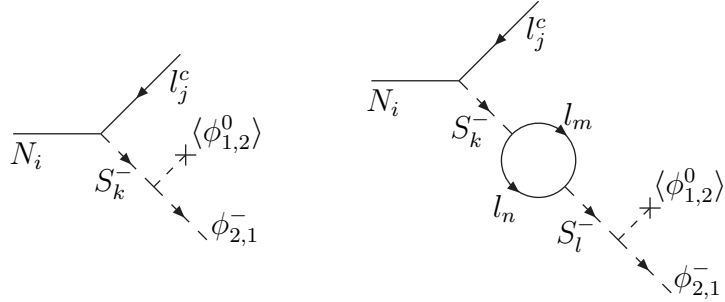
Taking into account the contribution 
of these two body decays to $\Gamma_{N_i}$ 
we have to add in Eq. (\ref{gam3bd}) 
the following term:
\begin{equation}
\Gamma^{(2)}_{N_{i}} 
\simeq \frac{1}{8 \pi} \sum_{jkl} h^k_{Rij} h^{l\ast}_{Rij}
 \frac{M_{N_i} v^2 \lambda_S^{k \ast} \lambda_S^l}{m^2_{S_k} m^2_{S_l}} \,,
\label{gam3bdbis}
\end{equation}
and, for the $\varepsilon_{N_i}$ asymmetries,
$A_{N_i}$ in Eq.~(\ref{eps3bd}) becomes now:
\begin{equation}
A_{N_i} \simeq
\frac{1}{\Gamma_{N_i}}
\frac{1}{(2 \pi)^3 } \frac{1}{12} \frac{\pi}{(4 \pi)^2}
\frac{M_{N_i}^5}{m^2_{S_1} m^2_{S_2}} 
+\frac{1}{\Gamma_{N_i}}
\frac{1}{(4\pi)^2}
\frac{M_{N_i}m^2_{h^+} v^2}{m^2_{S_1} m^2_{S_2}} \, .
\label{eps3bdAbis}
\end{equation}
with $m_{h^+}$ the mass of the physical light charged scalar which 
we will take typically around 300-400 GeV.
These two body contributions to the decay width and the asymmetry 
have about the same magnitude than the three body decay contributions 
because as can be seen from Fig.~4 they involve the same couplings. 
With respect to the three body decays, the two body decays are
enhanced by a smaller phase space suppression 
but are suppressed by $v^2/M_{N_i}^2$ and $m_{h^+}^2/M_{N_i}^2$ factors, 
so that
they have about the same order of magnitude.

Collecting all these results, to satisfy the out-of-equilibrium condition 
on the decay width, we 
need (for example with $M_N \sim 3$~TeV and $M_{S_{1,2}} \sim
10$~TeV):
\begin{eqnarray} 
|h^k_{Lij}| &<& 4 \times 10^{-7} \label{ooec13bd} \, , \\
|h^k_{Rij} f^{l\ast}_{mn}|&<& 1 \times 10^{-4} \, , \label{ooec23bd} \\
|h^k_{Rij}\lambda^{l \ast}_S|&<& 2 \times 10^{-4} \, \hbox{TeV}
\label{ooec33bd}
 \, .
\end{eqnarray}
Comparing Eqs.~(\ref{eps3bd}) and (\ref{eps3bdAbis}) with 
Eqs.~(\ref{ooec13bd})-(\ref{ooec33bd}),
it is not difficult to see that very large values of $\varepsilon_N$
(as large as $10^{-2}$) can be obtained with values of the 
couplings satisfying Eqs.~(\ref{ooec13bd})-(\ref{ooec33bd}).
However, as discussed below 
some care has to be taken with the potentially large damping effect
of various scatterings.

%%%%%%%%%%%%%%%%%%%%%%%%%%%%%%%%%%%%%%%%%%%%%%%%%%%%%%%%%%%%%%%%%%%%%%
\subsection{Scattering effects}
%%%%%%%%%%%%%%%%%%%%%%%%%%%%%%%%%%%%

Specially important are the scatterings with an intermediate scalar singlet:
\begin{itemize}
\item First the very fast L-violating scatterings proportional to
$\lambda_S^2 f^2$ (such as $l_L + l_L \leftrightarrow 
S \leftrightarrow \phi + \phi$)
can wash out very strongly the asymmetry. 
However, and this is an 
important point, the bounds on $\lambda_S^2 f^2$ we usually
obtain by requiring that these scatterings don't erase any preexisting 
asymmetry (at $T \sim m_{S_{1,2}}$) don't apply here. Much 
larger couplings can be taken 
here without large wash-out of the asymmetry because the asymmetry is produced
at $T \sim M_N$ which is below $m_{S_{1,2}}$. The large scattering 
effects at $T\sim m_{S_{1,2}}$ come mostly
from the $S_{1,2}$ resonance region.
For 
$T \sim M_N < m_{S_{1,2}}$ these contributions are significantly 
Boltzmann suppressed 
in Eq.~(\ref{gamscatt}).\footnote{To have a large enough 
Boltzmann suppression we
typically need that $m_S$ is larger than $m_N$ by a factor from $\sim 3$ to
15 depending on the values of the parameters.}
\item Second scatterings proportional to $h_R^4$, such as $N + N
\leftrightarrow l_R + \bar{l}_R$, which conserve lepton number but
bring the $N$ to thermal equilibrium if they are too fast. 
These scatterings put an upper bound on the $h_R$ couplings 
under which we will remain. The dependence of this bound in the values of
various other 
parameters is non-trivial. By taking $h_R$ below $10^{-3}$ we 
are in general safe.
\item Third the scatterings proportional to $h_R^2 f^2$
(such as $N + l_R \leftrightarrow S \leftrightarrow l_L + l_L$) 
or $h_R^2 \lambda_S^2$ (such
as $N+l_R \leftrightarrow S \leftrightarrow \phi + \phi$) which both
violate lepton number and can render 
the $N$ in thermal equilibrium down to temperature below its mass.
They are more dangerous than 
the $N + N \leftrightarrow l_R + \bar{l}_R$ scatterings
because their effect is enhanced by the charged scalar
resonance region contribution.
However this resonant region contribution
is also significantly Boltzmann suppressed at temperature below
the charged scalar mass when the asymmetry is produced. 
\end{itemize}
Other scatterings involving the large top Yukawa 
coupling (such as $N +l \leftrightarrow \phi \leftrightarrow \bar{t} + b$) 
have in general a small effect.
A set of values for which all these scatterings have moderate effects
when integrating the Boltzmann equations\footnote{This and other details 
will be given in a subsequent publication \cite{th}.} and which gives rise
to $n_L/s$ of order $10^{-10}$ at $T \sim 100$-$200$~GeV 
is for example the following:
\begin{eqnarray}
&& M_N= 4~\hbox{TeV}, ~~~~ m_{S_1} \sim 25~\hbox{TeV}, ~~~ m_{S_2} \sim 
30~\hbox{TeV}, ~~~  
f \sim 2 \cdot 10^{-2}, 
\nonumber \\
&& \lambda_S \sim 50 ~ \hbox{GeV}, ~~~~ h_R \sim 1.5 \cdot 10^{-4}, ~~ 
h_L \sim 10^{-8} \,. 
\label{set2}
\end{eqnarray}
Here for simplicity to obtain this result we took only the decay of
one right-handed neutrino assuming the other ones are heavier. We also took
all $f_{ij}^k$, $h_{Rij}^k$, $h_{Lij}^k$, $\lambda_S^k$ couplings with 
a same value $f$, $h_R$, $h_L$, $\lambda_S$ respectively. In
Eq.~(\ref{eps3bd}) maximal and positive CP-violating phases were taken.

%%%%%%%%%%%%%%%%%%%%%%%%%%%%%%%%%%%%%%%%%%%%%%%%%%%%%%%%%%%%%
\subsection{Neutrino masses}
%%%%%%%%%%%%%%%%%%%%%%%%%%%%%%%%%%%%%%%

In this model the neutrino masses can come either from the usual 
radiative effects in the Zee model sector of the model or from the
usual Yukawa couplings of the right-handed neutrinos via the seesaw mechanism.
Particularly interesting is the possibility that the neutrino masses 
in this model could be mostly due to the 
radiative Zee contribution (Fig.~5) which gives:
\begin{equation}
(m_\nu)_{ij}=\sum_k \frac{\lambda_S^k }{m_{S_k}^2-m^2_{h^+}} \frac{v_2}{v_1}
\frac{1}{(4 \pi)^2} \hbox{ln}\frac{m_{S_k}^2}{m_{h^+}^2} f^k_{ij}
(m^2_{l_j}-m^2_{l_i}) \, ,
\label{mnuZee}
\end{equation}
where $v_{1,2}$ are the vacuum expectation values
of both neutral Higgs bosons (for which we took $v_2 \sim v_1$). In 
fact interestingly it turns out that
the leptogenesis constraints on the various couplings (i.e. to have an asymmetry
of order $10^{-10}$) lead naturally to a value of the largest neutrino mass 
of the order of the 
SuperKamiokande bound ($\sim 0.1$~eV) from atmospheric neutrino data.
This is the case for example with the values of Eq.~(\ref{set2}).
To fit in addition the solar data, unlike in Eq.~(\ref{set2}), a 
hierarchy among the $f^k_{ij}$ couplings has to be assumed.
Assuming for simplicity $f_{ij}^1 \sim f_{ij}^2 \sim f_{ij}$, for example 
the LOW solution of solar neutrino
experimental data can be accommodated by assuming in 
Eq.~(\ref{mnuZee}) as in the Zee model the following 
hierarchy: $|f_{e \mu}|\sim 3 \cdot 10^2 |f_{e \tau}| \sim 10^{7}
|f_{\mu \tau}|$. A $\nu$ mass of order 0.1~eV requires in addition 
that $f_{e \mu}^k \lambda_S^k m^2_\mu/ m^2_{S_k}$ in Eq.~(\ref{mnuZee}) 
is of order $\sim 10^{-8}$~GeV. With such a value of this combination of
couplings
it turns out that the $l_\mu + l_\tau \leftrightarrow \phi + \phi$ scatterings
are fast enough to be in thermal equilibrium down to 
temperature below $M_N$. However it will
not erase all lepton asymmetries.
In fact it has been shown in Ref.~\cite{hhs} that for the LOW 
solution [as well as for
the vacuum oscillation (VO) solution]
all preexisting
lepton asymmetries will not be erased in the Zee model: the 
$f_{\mu \tau}$ coupling
is tiny enough to prevent the erasure of a preexisting 
$L_e-L_\mu-L_\tau\equiv L_1$ asymmetry. This quantum number will
be violated only by $l_\mu + l_\tau \leftrightarrow \phi + \phi$  
and $l_\mu +l_\tau \leftrightarrow l_e + l_\mu$ processes which
are slow enough to be out-of-equilibrium (in particular 
at $T \sim m_S$ and below).
%%%%%%%%%%%%%%%%%%%%%%%%%%%%%%%%%%%%%%%%%%%%%
%
\begin{figure}[!t]
\begin{center}
\begin{picture}(160,92)(0,0)
%1st diagram
\ArrowLine(0,25)(40,25)
\ArrowLine(80,25)(40,25)
\ArrowLine(120,25)(80,25)
\ArrowLine(160,25)(120,25)
\DashArrowArcn(80,25)(40,180,90){5}
\DashArrowArcn(80,25)(40,90,0){5}
\DashLine(80,65)(80,81){4}
\DashLine(80,25)(80,9){4}
\Line(77,78)(83,84)
\Line(77,84)(83,78)
\Line(77,6)(83,12)
\Line(77,12)(83,6)
\Text(20,15)[]{$\nu_{Li}$}
\Text(60,15)[]{$l_{Lj}$}
\Text(100,15)[]{$l_{Rj}$}
\Text(140,15)[]{$\nu_{Lj}$}
\Text(42,55)[]{$S^-_k$}
\Text(122,55)[]{$ \phi_1^-$}
\Text(80,-1)[]{$ \langle \phi^0_1 \rangle$}
\Text(80,92)[]{$ \langle \phi^0_2 \rangle$}
\end{picture}
\end{center}
\caption{One loop diagram inducing a Majorana neutrino mass. Here as in the
Zee model
we assumed that the leptons have Yukawa couplings 
only with one Higgs doublet
(i.e. with $H_1$).}
\label{fig5}
\end{figure}
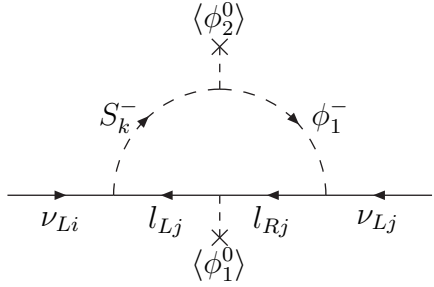
%%%%%%%%%%%%%%%%%%%%%%%%%%%%%%%%%%%%%%%%%%%%%
%
In our case, since we need these processes to be out-of-equilibrium only at
$T \sim m_N$ which is below $m_S$, these constraints are even less stringent
and it turns out that not only the LOW and VO solutions but also the large 
mixing angle (LMA) solution can be accommodated easily 
to create large asymmetries.
 
\newpage
With the LOW solution, for example the following set of values,
\begin{eqnarray}
&&M_N= 1~\hbox{TeV}, ~~~~ m_{S_1} \sim 4 ~\hbox{TeV},  ~~~ m_{S_2} \sim
5~ \hbox{TeV}, ~~~
f_{e \mu} \sim 2 \cdot 10^{-1}, 
\nonumber \\
&& \lambda_S \sim 50 ~\hbox{GeV}, ~~~~ h_R  \sim 5 \cdot 10^{-5}, ~~~ 
h_L < 10^{-8}. 
\label{set3}
\end{eqnarray}
leads to a large $L_1$ asymmetry together 
with two neutrino masses of order 0.1 eV.\footnote{The neutrino 
mass hierarchy in this case is that the two
heaviest neutrinos have masses differing by the small 
solar $\Delta m^2_{solar}$ and
the lightest neutrino differ from the other two by the larger atmospheric
$\Delta m^2_{atmos}$.}
In this case, the $L_1$ asymmetry is not erased at all, neither by the
$l_\mu+l_\tau \leftrightarrow \phi + \phi$ scatterings,
nor by the $l_\mu + l_\tau \leftrightarrow l_e + l_\mu$ scatterings
(including their scalar singlet on-shell part which 
are Boltzmann suppressed). 
We obtain $n_{L_1}/s \sim 10^{-9}$ (at 
$T \sim 100-200$~GeV) taking
maximal CP-violating phases (and
$n_{L_1}/s \sim 10^{-10}$ if these 
phases are reduced by one order
of magnitude).\footnote{Note that we could naively believe that if the
$f_{\mu \tau}$ couplings are small the produced $L_1$ asymmetry will also
be small. 
However the situation is more subtle: this is not the case because 
in $\varepsilon_N$, $L_1$ is also violated in the contribution
proportional to the large $f_{e \mu}$ couplings
via the $h_R$ couplings.}
Note that in this case also a $L_2 \equiv L_\mu -L_e - L_\tau$
asymmetry will be created. This asymmetry will be erased significantly because
$f_{e \tau}$ is not tiny enough to avoid the associated damping effect, 
but still we get $n_{L_2}/s \sim 10^{-11}$. 
Note that for such a value of $f_{e \tau}$, the produced $L_2$ asymmetry
turns out to be rather sensitive to the exact 
value of $f_{e \tau}$.  
Other sets of values could lead to a
larger $L_2$ asymmetry.
The produced $L_3 \equiv L_\tau-L_e-L_\mu$ asymmetry on the other hand will be 
completely negligible because $f_{e \mu}$ is large.

The LMA solution requires 
$|f_{e \mu}/ f_{e \tau}| \sim m^2_\tau/m^2_\mu \sim 3 \cdot 10^{2}$ 
together with
$|f_{e \tau}/ f_{\mu \tau}| \sim \sqrt{2} \Delta m^2_{atmos}/ 
\Delta m^2_{solar} \sim  10^{2}$ (see e.g.~Refs.~\cite{kan,grim}).
Assuming for example the set of values of Eq.~(\ref{set3}), this gives
$f_{\mu \tau} \sim 3 \cdot 10^{-5}$ which is tiny enough to prevent 
any sizable erasure of the $L_1$ lepton asymmetry 
from $l_\mu + l_\tau \leftrightarrow \phi + \phi$ or
$l_\mu + l_\tau \leftrightarrow l_e + l_\mu$ scatterings as with the
LOW solution. We also get in this case
$n_{L_1}/s \sim 10^{-9}$  and $n_{L_2}/s \sim 10^{-11}$ 
for maximal CP-violating phases. 
Note that the LMA solution leads nevertheless 
to nearly bi-maximal mixing and $\nu_e$ survival probability which become to 
be disfavoured by the data (see e.g.~\cite{pena} and \cite{framp,gla}). There 
are however ways to
avoid this problem \cite{grim,framp,yasue}, in particular 
by allowing both Higgs doublets to
have Yukawa couplings with leptons, in case the second mixing angle
can be reduced \cite{grim}.

Note that in this model not only a lepton asymmetry is created, but 
any preexisting asymmetries (if there are any) will be practically 
erased, so that 
the physics of higher 
energies would be irrelevant for baryogenesis.
A preexisting $L_3$ asymmetry will be erased in the same way as the 
produced $L_3$ asymmetry mentioned above.
A preexisting $L_2$ or $L_1$
asymmetry will be erased to a large extend at $T \sim m_S$ by the effect of 
the $S$ mediated $N + l_{e,\mu,\tau} \leftrightarrow l_e +l_\mu$ and
the $N + l_{e,\mu,\tau} \leftrightarrow \phi + \phi$ 
scatterings as well as for $L_2$ by 
the $l_e + l_\tau \leftrightarrow \phi + \phi$ scatterings. 
These scatterings erase the $L_{1,2}$ preexisting asymmetries much more than
the subsequently produced $L_{1,2}$ asymmetries 
because around $T \sim m_S$, unlike 
at $T \sim M_N$ and below, they are not Boltzmann suppressed. The
damping factor turns out to be
very sensitive to the exact values of 
the $f$ and $h_R$ couplings but  
for example for
the set of values of Eq.~(\ref{set3}) we observe that 
these eventual preexisting asymmetries are erased by 
several orders of magnitude \cite{th}.

Note also that in this case, where the neutrino masses are induced in
the Zee radiative way, neither for leptogenesis nor for neutrino masses,
do we need
the $h_{Lij}$ couplings. These couplings could all vanish 
without changing anything for neutrino masses and
leptogenesis.\footnote{Note that 
if at tree level $h_L=0$, an effective $h_L$ coupling is induced
at the one loop level from the $h_R$, $f$ and the lepton Yukawa couplings.
With the values of Eq.~(\ref{set3}) we get $h_L$ below $10^{-9}$.} 
In this case all tree level couplings are relatively large.
This is a unique feature of 
this case that, 
unlike all other models of leptogenesis at low scale, no special mass
degeneracies
or large hierarchies between the couplings is needed (except 
for the flavor structure associated to the neutrino masses).

Note however that if the $h_L$ couplings are non-vanishing at tree level
there is also the possibility that they contribute non-negligibly to 
the neutrino masses. 
A neutrino mass of order $\sim 0.1$~eV couldn't be generated in this model
by the lighter 
right-handed neutrino
$N_1$, whose decay is responsible for the 
asymmetry\footnote{As explained in section 3 above that 
would imply that $K_N$ is of order 100
which would give an inverse decay
damping effect very difficult to compensate by an appropriate choice 
of the parameters.},
but it could come from the two heavier neutrinos $N_{2,3}$. 
This could even be an alternative solution to explain that the second 
mixing angle is not as maximal as expected in the Zee model but we will 
not enter into these details. Here we just assumed that 
the $h_L$ coupings are tiny enough for the seesaw contribution
to be negligible and focussed on the radiative contribution.

%%%%%%%%%%%%%%%%%%%%%%%%%%%%%%%%%%%%%%%%%%%%%%%%%%%%%%%%%%%%%%%%%%%%%
\subsection{Discussion}
%%%%%%%%%%%%%%%%%%%%%%%%%%%%%%%%%%%%%%%%%%%%%%%%%%%%

A very attractive property of this model is that it could be tested 
at future accelerators. Heavy charged scalars could be observed
through electromagnetic interactions (e.g. from the large 
$l + \bar{l} \rightarrow
S^+ + S^-$ or $q + \bar{q} \rightarrow S^+ + S^-$ cross 
sections \cite{kan}). A charged scalar pair can also
be produced from $e^+ e^-$ or $e^- e^-$ annihilation with an intermediate 
right-handed neutrino in the t-channel \cite{nt} which also allows to observe
indirectly the right-handed neutrinos. 
The charged scalar can then decay to two left-handed leptons, 
two scalars or a right-handed neutrino 
and a right-handed charged lepton. They can also interact with an electron
of the detector to produce a right-handed (virtual) neutrino.
Note that right-handed 
neutrinos can 
be observed through their relatively large coupling to a right-handed 
fermion and a charged scalar singlet. This has to be contrasted with 
leptogenesis 
models where the right-handed 
neutrinos couple only to left-handed doublet, in case they should have either a 
mass far beyond the reach of any foreseeable future 
accelerator or
tiny couplings, which makes them difficult to observe. To produce
right-handed neutrinos,
the $S \rightarrow l_R + N$ decay
and $l_R 
+ \bar{l}_L \rightarrow N + l_L$
scatterings (with a 
charged singlet scalar in the t-channel)
are particularly interesting here. With for example the values 
of Eq.~(\ref{set3}), the $S$-mediated cross section is 
larger than
the usual $Z$-mediated $e^+ + e^- \rightarrow \nu + N$
cross section \cite{rai} induced by neutrino mixing. Above the $S$ production
threshold the production of $N$ from $S \rightarrow l_R + N$ is completely 
dominant.
Once it is produced the $N$ will decay 
very slowly (but still inside the detector) or it will 
interact with an electron of the detector. Note also that a value
of $f_{e \mu}$ of order $10^{-1}$ as in Eq.~(\ref{set3}) is about one order
of magnitude smaller than the experimental upper bounds on it 
from electroweak data \cite{smir,mitu}. The values 
of the $h_R$ and $f$ couplings 
we have considered are in particular small enough not to violate the bound on
$\Gamma(\mu \rightarrow e \gamma)$ \cite{smir,mitu}. The contribution of the
$f$ couplings 
to the muon anomalous magnetic moment \cite{mitu} doesn't exceed
$a_\mu \sim 10^{-13}$, i.e.~is 3-4 orders of magnitude smaller
than the present experimental sensitivity.

In a more 
speculative vein note also that, if 
we add two additional "complex" neutral scalar singlets,
this model displays the puzzling "symmetry" that for
every left-handed doublet there are two associated "right-handed"
$SU(2)_L$ singlets (i.e.~for the scalars as well as for the leptons).
This could be due to the breaking of a right-symmetry in 
some left-right symmetric model. Once the right symmetry
is broken from every right doublet remain just two
singlets.

Before concluding note finally that in appendix B we comment on the
leptogenesis model at the TeV scale recently 
proposed in Ref.~\cite{nt}. We explain why we disagree with the statement
that phase space suppressed two body decay could lead to an enhancement of
the asymmetry. 
%%%%%%%%%%%%%%%%%%%%%%%%%%%%%%%%%%%%%%%%%%%%%%%%%%%%%%%%%%%%%%%%%%%%%%%%%%
\section{Summary}
%%%%%%%%%%%%%%%%%%%%%%%%%%

In summary after discussing the various issues associated with low scale
leptogenesis, and in particular the problems of existing neutrino mass
models, we showed how a leptogenesis mechanism based on 
three body decays could avoid easily these problems. In general 
three body decays
require models with a particle content which is more elaborate
than in models with two body decays. But in
contrast with the usual two body decays 
where the natural scale for producing a large asymmetry is  
around $\sim 10^{10}$~GeV (or above),
for three body decays the 1-10~TeV scale is a perfectly possible scale.
This does not require unnatural large coupling hierarchies
or mass degeneracies like the other two mechanisms with two body decays do.
Moreover the values of the couplings in the 
three body decays which are required 
for leptogenesis have typically the size required to induce
the correct neutrino masses radiatively. 
Three body decay induced leptogenesis and one loop induced neutrino masses
constitute therefore a TeV scale alternative to the 
usual leptogenesis and neutrino mass framework with masses
in the $10^{10}$-$10^{15}$~GeV range; in the later case the leptogenesis is
induced by
two body decays and the neutrino masses are induced in 
the seesaw mechanism.
This three body decay mechanism could be operative in many different contexts. 
We implemented it 
in a simple and minimal model with right-handed neutrinos and 
charged scalar singlets. Beyond the fact that this model leads to 
a large enough lepton asymmetry at low scale, it displays a number 
of attractive properties:
\begin{itemize}
\item Neutrino masses and mixings in agreement with the 
solar and atmospheric neutrino experimental data can be produced easily in
this model
through the radiative Zee mechanism. The radiative Zee contribution
to the largest neutrino mass
is naturally
of order the SuperKamiokande bound.
\item Instead of, as usually discussed in the literature, only 
erasing some or all possible preexisting lepton asymmetries,
the scalar singlet 
Zee couplings here can lead both: a large erasure of all preexisting
asymmetries and the subsequent creation of a new lepton asymmetry.
This is related to the fact that three body decays involve naturally 
two scales,
the mass of the decaying particle $M_N$ and
the mass of the virtual particle $m_S$. 
At $T\sim m_S$ the preexisting asymmetries 
are washed out and at $T \sim M_N$
a new asymmetry is created.
\item Beside the scalar singlet Zee couplings, 
the leptogenesis in this model is not based on the couplings
of the right-handed neutrinos to the left-handed leptons,
which we could take as vanishing, but on their couplings 
to the right-handed leptons.
These later couplings are relatively large, which could allow
to observe the right-handed neutrinos much more easily than in the usual 
Fukugita-Yanagida case
where the 
right-handed neutrinos couple only to left-handed leptons.
\item In this model the right-handed neutrinos 
could have a mass as low as $\sim 1$~TeV and the 
charged scalar singlets could have a
mass as low as a few TeV. 
\end{itemize}

The origin of the baryon asymmetry and of the related neutrino masses
could be therefore directly tested 
at future accelerators running around the 10 or 20 TeV scale.
%
%%\newpage
%
\renewcommand{\textfraction}{1.0}
%
%\newpage
\vspace{1.1cm}
\begin{center}{\large Acknowledgements}
\end{center}  
It is a pleasure to thank Ernest Ma and Utpal Sarkar
for many useful discussions. We 
also wish to thank
David Del\'epine and Jean-Marc G\'erard for useful discussions. This work was 
supported by the TMR, EC-contract No. ERBFMRX-CT980169 (EuroDa$\phi$ne).
%
%%%%%%%%%%%%%%%%%%%%%%%%%%%%%%%%%%%%%%%%%%%%%%%%%%%%%%%%%%%%%%%%%%%%%%

\vspace{0.9cm}
\begin{appendix}

\section{Boltzmann equations}

It is customary to express the number of particles "i" in term
of
$X_i=n_i/s$ which gives the number of particle "i" per 
comoving volume $n_i$ divided by the entropy per comoving volume:
\begin{equation}
s=g_\ast \frac{2 \pi^2}{45} T^{3}\label{entropy}.
\end{equation}
Similarly the total lepton number per comoving volume is
parametrized by $X_L=n_L/s=(n_l-n_{\bar{l}})/s$. As a 
function of $z=M/T$ with $M$ an arbitrary mass scale (for 
example $M=M_a$, the mass of the particle "a" 
whose decay is at the 
origin  of the asymmetry), $X_a(z)$ and $X_L(z)$ are 
given by \cite{kt}:
\begin{eqnarray}
\frac{dX_a}{d z}&=&
- z K_{a} \frac{K_1(z)}{K_2(z)}
\Big( \frac{M_a}{M}  \Big)^2
\Big({X_{a}}-{X^{eq}_{a}}  \Big)\nonumber\\
&&+ z  \frac{1}{s H(M)} \Delta n_a  
\frac{X_{i_1} X_{i_2} \cdots}{X_{i_1}^{eq} X_{i_2}^{eq} \cdots} 
\gamma^{eq}_{scatt}(i_1 + i_2 + \, \cdots 
\rightarrow f_1 + f_2 + \, \cdots )\, ,
\label{bol1}\\
\frac{d X_L}{d z}&=& \sum_{a,k}
z K_a   \frac{K_1(z)}{K_2(z)}
\Big( \frac{M_a}{M}  \Big)^2
\Big[\, \varepsilon_a ({X_a}-{X^{eq}_a} ) 
- \frac{1}{2}   \frac{X^{eq}_a}{X_l^{eq}} X_L \Big] \nonumber\\
&& +  \frac{z}{sH(M)} \Big( \Delta n_l
\frac{X_{i_1} X_{i_2} \cdots }{X_{i_1}^{eq} X_{i_2}^{eq} \cdots }
\gamma^{eq}_{scatt}(i_1 + i_2 + \, \cdots \rightarrow f_1 + f_2 + \, \cdots )
\nonumber \\
&&- \Delta n_{\bar{l}}
\frac{X_{i_1} X_{i_2} \cdots }{X_{i_1}^{eq} X_{i_2}^{eq} \cdots } 
\gamma^{eq}_{scatt}(i_1 + i_2 + \, \cdots \rightarrow f_1 + f_2 + \, \cdots )
\Big) \,
.\label{bol2}
\end{eqnarray}
In those equations $K_{1,2}$ are the usual modified Bessel functions and
$X_i^{eq}(z)=n_i^{eq}(z)/s$ gives the number density of
the particle "i" in thermal equilibrium.
It is a good approximation to use Maxwell-Boltzmann statistics:
\begin{equation}
n^{eq}_{i}=g_{i}\frac{M_{i}^2}{2 \pi^2}
T K_2(M_{i}/T) \, ,
\label{maxbol}
\end{equation}
with $g_i$ the number of degree of freedom of the particle "i". For 
massless particle the number density is:
\begin{equation}
n^{eq}_i=\frac{g_i T^3}{\pi^2}
\label{ngam}
\end{equation}
with for the photon $g_\gamma=2$.
In Eqs.~(\ref{bol1})-(\ref{bol2}), $K_{a}\equiv \Gamma_{a}/H(M_{a})$ 
parametrizes the effect of the decays and 
inverse decays of $N_i$.
If the 
out-of-equilibrium decay condition 
of Eq.~(\ref{ooecfy}) is not satisfied, the inverse decay term 
of Eqs.~(\ref{bol1})-(\ref{bol2}) (i.e. 
the terms proportional to $K_a X_a^{eq}$)
will compensate
the decay term (i.e. the terms proportional to $K_a X_a$). In this 
way the decaying particle remains 
at equilibrium and no asymmetry is produced.  
In Eqs.~(\ref{bol1})-(\ref{bol2}),
the $\gamma^{eq}_{scatt}$ are the scattering reaction densities
which can be obtained from the scattering cross sections
in the following way:
\begin{equation}
\gamma^{eq}_{scatt}= \frac{T}{64 \pi^4} \int_{s_0}^{\infty}
ds\, \hat{\sigma}(s)\, \sqrt{s} \, K_1(\sqrt{s}/T) \, ,
\label{gamscatt}
\end{equation}
where $\hat{\sigma}(s)$ is the reduced cross section and is given
by $2 [s-(m_1+m_2)^2][s-(m_1-m_2)^2] \sigma(s) /s$ with $\sigma(s)$ the
cross section. 
$m_{1,2}$ are the masses of the particles in the 
scattering initial 
state and $s_0$ is the scattering threshold.
$\Delta {n_i}$ is 
the net number of particle "i" which were 
created in one scattering $i_1 + i_2 + \,
\cdots \rightarrow f_1 + f_2 + \, \cdots$.
The
scatterings can damp the asymmetry in two different ways. First if they don't 
conserve lepton number
they are present in Eq.~(\ref{bol2}) and, if fast enough, can directly 
reequilibrate the lepton number to 0. Secondly 
if they change the number of particle "a" 
they are present in 
Eq.~(\ref{bol1}) and if 
fast enough they will damp the asymmetry by imposing 
the species "a" to remain in equilibrium with the thermal bath
down to temperature much below its mass.
These scattering effects in Eq.~(\ref{bol1}) appear to be especially 
important
in most of the possible leptogenesis model candidates at the TeV scale.

\section{Comments on the model of Ref.~\cite{nt}}

The model recently proposed in Ref.~\cite{nt} is based on a particle 
content which is similar to the one proposed in section 5.
However quite different assumptions on the various possible 
couplings and on the leptogenesis mechanism have been made.
It is claimed that, if the Fukugita-Yanagida model cannot lead easily to 
a large enough asymmetry, the self-energy diagram of Fig.~1 above 
with $h_R$ couplings in the loop instead of $h_L$ couplings (i.e. with 
a $l_R$ and a charged scalar singlet $S$ in the 
loop instead of a $l_L$ and a $\phi$),
can lead to a much larger asymmetry by taking large $h_R$ couplings.
It is also said that this requires a large phase space suppression
of the S decay width to $l_R$ and $N$ (a value of $y=1-(m_S^2/m_N)^2$
as small as $10^{-4}$-$10^{-5}$ has been assumed) to avoid associated wash out
of the asymmetry. 
%However it must be noted
%that
%if $S$ decays to $l_R$ and $N$ the asymmetry vanishes because the self energy
%doesn't give rise to any imaginary part. In addition, if now it 
%is $N$ which decays to $l_R$ and $S$,
Note however that in this case the asymmetry is proportional 
to $(1/8 \pi)\cdot (h_L^2 h_R^2 y^2)/(h_L^2+h_R^2 y^2)$ (assuming 
universal $h_L$
and $h_R$ couplings for all right-handed neutrinos as it has been done).
From this result we agree that the numerator can be enhanced only if 
$h_R y$ is taken large since $h_L$ is anyway constrained to be small.
By unitarity this increase of the numerator 
will be nevertheless compensated by a same increase of 
the denominator. Therefore with respect to the Fukugita-Yanagida model
we don't gain anything by taking $h_R$ larger than $h_L$ in this
self-energy diagram (and this for any value of
$y$).
The large asymmetry obtained in Ref.~\cite{nt} has been obtained from a 
large right-handed neutrino mass degeneracy (as can be done 
in the Fukugita-Yanagida
model) and from omitting the $h_R^2 y^2$ term in the $N$ decay width, hence in
the denominator of the asymmetry. To take into account this term
in the decay width implies a decrease of the asymmetry and 
in addition an enhancement of the damping factor
from inverse decay processes.
Note also that the $h_R$ couplings have 
been assumed in Ref.~\cite{nt} to be of order
$\sim 1$. This would induce an additional large damping effect from 
the $N+N \leftrightarrow l_R + \bar{l}_R$ scatterings
which has not been taken into account.
The suppression factor due to these scatterings 
will not be as large as $\sim 10^6$ as in the 
example given in section 3 above because $K_N\sim 100$ and $m_N \sim 10$~TeV 
have been taken here (instead of $K_S \sim 1$ and $m_S \sim 1$~TeV in the
example of section 3)  
but still we estimate it to be
at least of order $10^2$-$10^3$. Taking into account 
these various suppression 
effects, it would be
very difficult to induce a large enough asymmetry in the 
way of Ref.~\cite{nt}, unless we assume
a huge degeneracy of the right-handed neutrinos as can be done in the 
usual Fukugita-Yanagida model.
\end{appendix}
%\vspace{1.2cm}
%
%%%%%%%%%%%%%%%%%%%%%%%%%%%%%%%%%%%%%%%%%%%%%%%%%%%%%%%%%%%%%%%%%%%%%%%

%
\newpage
\renewcommand{\textfraction}{1.0}
\end{document}